\documentclass[%
aip,jcp,amsmath,amssymb, reprint, twocolumn, nofootinbib
]{revtex4-2}
\usepackage{amsmath}
\usepackage{mathtools}
\usepackage{float}
\usepackage{amsfonts}
\usepackage{amssymb}
\usepackage{dcolumn} 
\usepackage{array}

\newcolumntype{P}[1]{>{\centering\arraybackslash}p{#1}}
\newcolumntype{M}[1]{>{\centering\arraybackslash}m{#1}}
\newcolumntype{C}[1]{>{\centering\arraybackslwash}p{#1}}
\usepackage{float}
\usepackage{psfrag}
\usepackage{tabularx}
\usepackage{stackengine}
\usepackage{amssymb}
\usepackage{mathtools}  
\usepackage{xfrac} 
\usepackage[T1]{fontenc}
\usepackage{graphicx}
\usepackage{diagbox}
\usepackage[caption=false]{subfig}
\usepackage{tikz}
\usetikzlibrary{positioning}
\usetikzlibrary{arrows}
\usetikzlibrary{trees}
\usepackage{amssymb}
\usetikzlibrary{decorations.pathmorphing}
\usetikzlibrary{decorations.markings}
\usetikzlibrary{automata,positioning}
\usepackage{braket}
\usepackage{rotating}
\usepackage{adjustbox}
\usepackage{ragged2e}
\usepackage{simplewick}
\usepackage{simpler-wick}

\usepackage[]{lineno}

\usepackage{csquotes}
\usepackage{physics,amsmath}
\usepackage{xcolor}
\usepackage{fancyhdr}
\UseRawInputEncoding
\usepackage[normalem]{ulem}

\usepackage{scalerel}
\usepackage{hyperref}
\usepackage{appendix}

\nolinenumbers
\hypersetup{
  colorlinks   = true, 
  urlcolor     = blue, 
  linkcolor    = blue, 
  citecolor   = red 
}
\begin{document}

\author{Srushti Patil}
\thanks{These two authors contributed equally.}
\affiliation{ NNF Quantum Computing Programme, \\ Niels Bohr Institute,
University of Copenhagen, Jagtvej \\ 155 A, Copenhagen N DK, 2200 \\ Copenhagen, Denmark}

\author{Dibyendu Mondal}
\thanks{These two authors contributed equally.}
\affiliation{ Department of Chemistry,  \\ Indian Institute of Technology Bombay, \\ Powai, Mumbai 400076, India}

\author{Rahul Maitra}
\email{rmaitra@chem.iitb.ac.in}
\affiliation{ Department of Chemistry,  \\ Indian Institute of Technology Bombay, \\ Powai, Mumbai 400076, India}
\affiliation{Centre of Excellence in Quantum Information, Computing, Science \& Technology, \\ Indian Institute of Technology Bombay, \\ Powai, Mumbai 400076, India}

\title{Machine Learning Approach towards Quantum Error Mitigation for Accurate Molecular Energetics}

\begin{abstract}
    Despite significant efforts, the realization of the hybrid quantum-classical algorithms has predominantly been confined to proof-of-principles, mainly due to the hardware noise. With fault-tolerant implementation being a long-term goal, going beyond small molecules with existing error mitigation (EM) techniques with current noisy intermediate scale quantum (NISQ) devices has been a challenge.  That being said, statistical learning methods are promising approaches to learning the noise and its subsequent mitigation. We devise a graph neural network and regression-based machine learning (ML) architecture for practical realization of EM techniques for molecular Hamiltonian 
    without the requirement of the exponential overhead. Given the 
    short coherence time of the quantum hardware, the ML model is trained with either 
    ideal or mitigated expectation values over a judiciously chosen ensemble of shallow
    sub-circuits adhering to the native hardware architecture. The hardware connectivity 
    network is mapped to a directed graph which encodes the information of the native
    gate noise profile to generate the features for the neural network. The training 
    data is generated on-the-fly during ansatz construction thus removing the computational overhead. We demonstrate orders of magnitude improvements in predicted energy over
    a few strongly correlated molecules. 
\end{abstract}


\maketitle
\section{introduction} \label{sec: intro}

Finding the exact ground state of general many-body systems is computationally hard and belongs to classically intractable problems\cite{expo,expo1}. In fact, finding the ground state of a $k$-local Hamiltonian is QMA-complete \cite{PRXQuantum.3.020322}. Nevertheless, with recent advancements in developing Noisy Intermediate-Scale Quantum (NISQ) devices\cite{quantum_pro,Trapped_ion}, handling the exponential scaling of resources has become tractable. Even though state-of-the-art quantum computers do not offer substantial quantum advantages due to quantum decoherence and a limited number of qubits, there has been active research in developing quantum algorithms and error mitigation techniques to harness the potential of quantum computing which might even allow us to study emerging phenomena in physics and chemistry.


Variational Quantum Eigensolver  (VQE)\cite{peruzzo2014variational} is a promising near-term 
application to find the ground state energy as it is designed to run on
noisy machines with low coherence time \cite{kandala2017hardware}. VQE takes the parameterized quantum \textit{ansatz} 
$\ket{\Psi(\theta)}$, and classically optimizes parameters to reach the ground state of a given molecular Hamiltonian.
\begin{equation}
    E(\theta)=\bra{\Psi(\theta)}\hat{H}\ket{\Psi(\theta)}
    \label{vqe}
\end{equation}
There has been an extensive study on the choice of ansatz $\ket{\Psi(\theta)}$\cite{ucc_review,Tilly_2022}. Within the domain of 
chemistry the unitary coupled cluster with singles and doubles (UCCSD)\cite{ucc_review,Tilly_2022,ducc} ansatz has emerged as a prevalent choice for the trial wavefunction where the unitary is expressed as:
\begin{equation}
     U(\theta)=e^{\tau_1(\theta_1)+\tau_2(\theta_2)} ;\hspace{0.3cm} \tau(\theta) = T(\theta)-T^{\dagger}(\theta)
\end{equation}
Here, $T(\theta)=\theta_{i,j,...}^{a,b,...}a^{\dagger}_aa^{\dagger}_b...a_ja_i$ is the 
excitation operator where $i,j,...$ are the occupied orbital indices and $a,b,...$ are the 
virtual orbital indices with respect to reference Hartree-Fock (HF) state.
In our implementations, we've reduced excitation operators by focusing solely on the dominant ones in the unitary, as detailed in section \ref{sec: theory}. Although VQE has been proven to be resilient to noise for smaller quantum systems,  inherent noise present in the current quantum processors deteriorates the performance of VQE for larger many-body Hamiltonian and complex quantum circuits. The sampling complexity of the expectation value estimation within VQE scales as $O(\frac{1}{\epsilon^2})$ for given precision $\epsilon$, leading to exponential runtime and thus preventing it from going beyond proof of principles\cite{PhysRevA.92.042303}. The theory of quantum fault tolerance could be used to deal with noise in the long run; however, in the current NISQ  era, focusing on mitigating the noise instead of eliminating it is practically more realistic.  
The most captivating feature of Quantum Error Mitigation (QEM) techniques 
is their ability to minimize the noise-induced distortion in expectation 
values calculated on noisy hardware. Current methods such as Zero Noise 
Extrapolation (ZNE)\cite{temme2017error}, Dynamical Decoupling 
 \cite{PhysRevLett.95.180501,DD1}, Probabilistic Error 
Cancellation \cite{van2023probabilistic}, etc rely on modifying the existing circuit into a logically 
equivalent circuit but with an increased number of gates and hence gate 
errors. The 
requirement of such additional mitigating circuits and exponential 
overheads make these methods impractical. Along with VQE, works have 
been done in the Projective Quantum Eigensolver (PQE)\cite{PQE} 
framework using ZNE \cite{10.1063/5.0166433}.

Statistical learning methods are known to learn complex functions underlying the input data very well and the field of QEM is no exception to such models, as they eliminate the need for extra mitigating circuit execution, reducing runtime overheads. Although initial machine learning (ML) exploration for QEM has shown promising results, \cite{9226505, bennewitz2022neural, kim2022quantum, saravanan2022graph, liao2023machine, Czarnik2021errormitigation} they are limited
to shallow circuits and small fermionic systems. Scalable and cost-efficient QEM techniques must be tested for molecular
Hamiltonians for strongly correlated systems beyond two-electron as 
they pose the most versatile and diverse circuit classes. However, existing schemes face information-theoretic limitations, such as entanglement spreading. To achieve quantum advantage, entanglement is necessary, but it comes at the cost of rapid noise spreading. Even when operating at 
$poly (log (log(n)))$ depth, estimating the expectation values of noiseless
observables, in the worst-case scenario, requires a super-polynomial 
number of samples\cite{quek2023exponentially}. Many QEM strategies like ZNE 
use exponentially many samples with respect to circuit depth, implying 
that the information should quickly get degraded due to sequential 
noise effects, requiring exponential overhead to remove the accumulated error \cite{PhysRevLett.131.210602}. Hence such QEM strategy is primarily
limited to model Hamiltonian and very shallow quantum circuits.

In this work, we develop GraphNetMitigator (GNM) - a combined Graph Neural Network (GNN) 
and regression-based learning model to mitigate the noisy expectation values obtained through VQE for strongly correlated 
molecular systems over their entire potential energy profile across diverse quantum devices. We use snippets of the entire circuit
consisting of single- and two-body fermionic excitations for model
training and use them to predict the mitigated expectation values 
for the complete ansatz. The use of GNN is motivated from the fact 
that graph representations are the best modeling choice for such type 
of data where the instances have a similar circuit structures but 
differ in terms of two-qubit gate connectivity. We dynamically 
generate a portion of noisy training data during ansatz construction 
to ensure runtime efficiency. To validate the robustness of the GNM, we train it in two different settings: (1) assuming access to the ideal quantum computer 
(simulator-generated expectation values are used as labels) and (2) assuming we 
do not have access to the fault-tolerant quantum computer. In the latter case, 
we generate the labels by sequentially mitigating quantum error (see 
supplementary materials for details) which was found to produce accurate 
energetics for shallow quantum circuits without additional mitigation
circuits. We employed CNOT-efficient implementation throughout to
reduce circuit depth drastically \cite{Yordanov_2020,cnot-effi}. The 
performance of the GNM is tested on linear $H_4$ 
and $BH$ molecules across their dissociation energy surface 
with diverse noise profiles derived from real hardware. We also 
study the robustness of GNM over other QEM techniques with varied 
synthetic noise strengths.

\section{Theory} 

The current limitations of the hardware has motivated
research to construct shallow dynamic ansatz that can be realized in
NISQ hardware\cite{ADAPT,uccgsd}. In often cases, the construction of the dynamic ansatz 
relies on the evaluation of gradients\cite{ADAPT} in quantum hardware and under
noisy scenario, such ansatz are often sub-optimally expressive. We 
previously advocated the importance of minimizing (or entirely 
bypassing) the usage of quantum resources while the dynamic ansatze 
are constructed such that the resulting ansatze maintain the requisite
expressibility\cite{Compass,compact,rbm}. Of course, their execution (through 
whatever method they are generated) in a quantum device requires additional EM 
routines for accurate prediction of energetics. In this manuscript, 
while our principal focus is to develop the ML based EM strategy
for any arbitrary disentangled ansatz\cite{ducc}, we also propose a robust
methodology that forms a dynamic ansatz \textit{in a noisy quantum architecture} which is nearly identical to an ansatz constructed 
under noiseless environment from the same operator pool. 
Such an undertaking serves twin 
desirable goals: firstly, it automatically generates the features 
needed for the regressor of our subsequent ML model and more 
importantly, due to its compatibility with a sequential reference 
error mitigation (SREM) strategy that generates 
mitigated results (which are used as training labels) for sub-circuits
on the fly as the ansatz gets constructed. In short, this entire
workflow can be executed in noisy quantum device without any 
implicit or explicit dependence to the fault-tolerant qubits. In the following section,
we begin with the dynamic ansatz construction strategy in a noisy 
environment and demonstrate how this strategy generates additional
features that are used up by the ML regressor.

\subsection{Construction of a Dynamic Ansatz in Noisy Architecture}\label{sec: theory}

For all the two-body excitations characterized by the hole-particle composite index
tuple $I$, we calculate the one-parameter energy functional by optimizing the single 
parameter circuits 
\begin{equation}
    E_I =\bra{\phi_{HF}}e^{-\tau_I(\theta_I)}\hat{H}e^{\tau_I(\theta_I)}\ket{\phi_{HF}} \forall I
\end{equation}
where $\theta_I$ is the sole variational parameter in each such optimization.
Furthermore, we calculate the "noisy" reference energy for each of 
these one-parameter circuits with the parameter values set to be zero:
\begin{equation}
     E_I^0 =\bra{\phi_{HF}}e^{-\tau_I(\theta_I=0)}\hat{H}e^{\tau_I(\theta_I=0)}\ket{\phi_{HF}} \forall I
\end{equation}
With the knowledge of the noisy reference energy for each 
one parameter circuit, we only screen in those 
operators in our ansatz for which $E_I^0 - E_I > \epsilon$,
where $\epsilon$ is a suitably chosen threshold\cite{Compass,energy_sorting}. With $N_T$ 
number of such parameters screened in, we align them in descending order of their 
stabilization energy to form the ansatz:
\begin{equation}
    U_2 = ...e^{\tau_L}...e^{\tau_K}...e^{\tau_J}...
    \label{eq: U}
\end{equation}
where $|E_J|...>..|E_K|..>..|E_L|...$ This
forms the basis of our ansatz towards the development of the GNM mitigation 
model. The determination of 
$E_I$ for various $I$ can be performed independently in
parallel. Note that the set of noisy expectation value $E_J$'s are to be 
used as one of the features of the ML model (\textit{vide infra}). Also it 
is pertinent to mention that
in our actual implementation, we used a fixed seed 
value of the sampling error for reproducibility which is compatible with the modern 
hardware noise suppression techniques.

The method outlined above selects only effective double excitation operators. The 
dominant single excitation operators are chosen based on spin-orbital 
symmetry considerations. Only those single excitation operators are selected for which the 
product of the irreducible representation of the occupied and virtual spin-orbital is 
totally symmetric ($A$) and thus their selection does not
incur any additional quantum resource utilization\cite{PRA_2022,SSSA-vqe}. If 
there is a $M$ number of such totally symmetric single 
excitation operators then the final unitary can be 
represented as:
\begin{equation}\label{final_ansatz}
    U = \prod_{a=1}^{M}e^{\tau_1^a}  U_2
\end{equation}
It is important to note the ordering of the operators 
in which they appear in the ansatz Eq. \ref{final_ansatz} as the various circuit 
snippets (as required for training the NN) uses this ordered pool of operators.

\begin{figure*}
    \centering
    \includegraphics[width= 17cm, height=5cm]{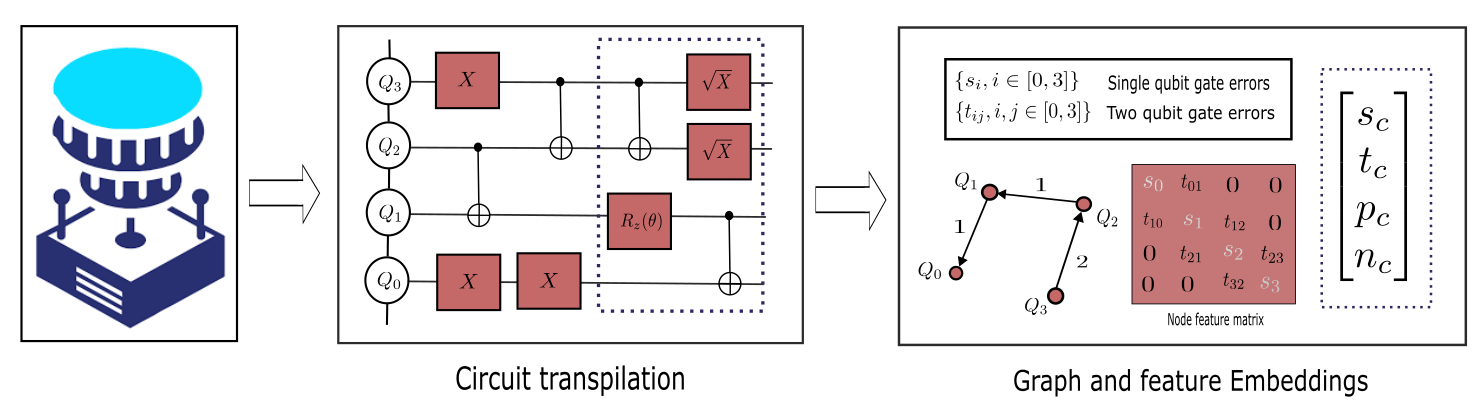}
    \caption{\textbf{Graph embedding of a demonstrative 4-qubit circuit transpiled on quantum hardware with native gates and linear qubit connectivity. CNOT connectivity of the transpiled circuit adhering to the device's qubit connectivity is encoded in a directed graph. Single and two-qubit gate errors ($s_i$ and $t_{ij}$) of the device are encoded in the node feature matrix. The rest of the features used while training; $s_c$ ($t_c$): single (two) qubit gate count, $p_c$: number of parameters, $n_c$: noisy expectation value calculated using VQE on the noisy device are derived from the circuit.}}
    \label{fig: gnn embd}
\end{figure*}

\subsection{Additional Feature Vector and Label Generation for Data Training} \label{sec: tr data}


For circuit generation that forms our training data, we consider a predefined operator pool 
forming the ansatz $U$ given by Eq. \ref{final_ansatz}. The circuits over which the data training 
is performed are snippets of the final parametrized quantum 
circuit corresponding to the ansatz given by 
Eq. \ref{final_ansatz}. There are various possible approaches one may adopt to select these 
smaller circuits (to be referred interchangeably as \textit{snippet circuit}). 
An option is to randomly group the operators from the pool of selected operators. However, 
we specifically choose to generate and mitigate the noisy data with up to 3-parameter 
circuits and the corresponding mitigated energy values of 
these snippet circuits form the training labels 
for the GNM. Towards this, we generate the snippets 
of all allowed 1-parameter circuits
\{($e^{\tau_J}$), ($e^{\tau_K}$),... ($e^{\tau_L}$)\}, 
ordered 2-parameter circuits 
\{($e^{\tau_K}e^{\tau_J}$), ($e^{\tau_L}e^{\tau_K}$),... ($e^{\tau_L}e^{\tau_J}$)\}, 
and ordered 3-parameter circuits \{($e^{\tau_L}e^{\tau_K}e^{\tau_J}$)\}, taken from the 
ordered set of operators selected in the parametrized quantum circuit.
The ordering of the operators is preserved in these snippet (training) 
circuits as they 
appear in Eq. \ref{final_ansatz}. The corresponding quantum circuits 
are transpiled for the chosen hardware as per its qubit connectivity 
and native gate configuration. Each circuit constructed adhering to the 
above ordering and hardware layout has a one-to-one 
correspondence with a directed graph that captures all the 
CNOT connections and acts as an input to our GNN of hybrid 
GNM. Details of graph construction are given in 
section \ref{sec: graph form}. We derive the additional features 
like the \textit{noisy expectation values, number of two-qubit 
gates, number of single-qubit gates, and the number of parameters} 
from the transpiled circuit and they serve as inputs to our regressor. 
As mentioned before, our model is trained in two different 
ways; (1) with ideal simulated values and (2) with SREM values.
In SREM, the noisy expectation values from the transpiled snippets are 
sequentially mitigated starting from the 1-parameter snippets to
all the way down to the 3-parameter snippets (see supplementary
materials for their detailed working equations). 
 These mitigated 
energy values are used as labels of GNM for training.
This is also to be noted that we could have used any
mitigation strategy for label generation but the accompanying 
exponential overhead of such strategies (like ZNE and PEC) makes 
them difficult to be cost-efficient and thus we rely upon the
sequential error mitigation strategy for the label generation. 

In principle, the SREM values generated from the 1, 2, 3 parameter snippets for the labels are nearly as accurate as when the model is trained against the ideal values.
To summarize, \textit{our approach, in principle, does 
not require any fault-tolerant qubit, nor does it necessarily 
have any dependence on any existing error mitigation strategy 
that itself requires exponential quantum overheads to generate 
the training labels}. In the next subsections, we discuss the 
details of the graph formulation and the subsequent GraphNetMitigator Architecture.

\begin{figure*}
    \centering
    \includegraphics[width= 17.5cm, height=9.5cm]{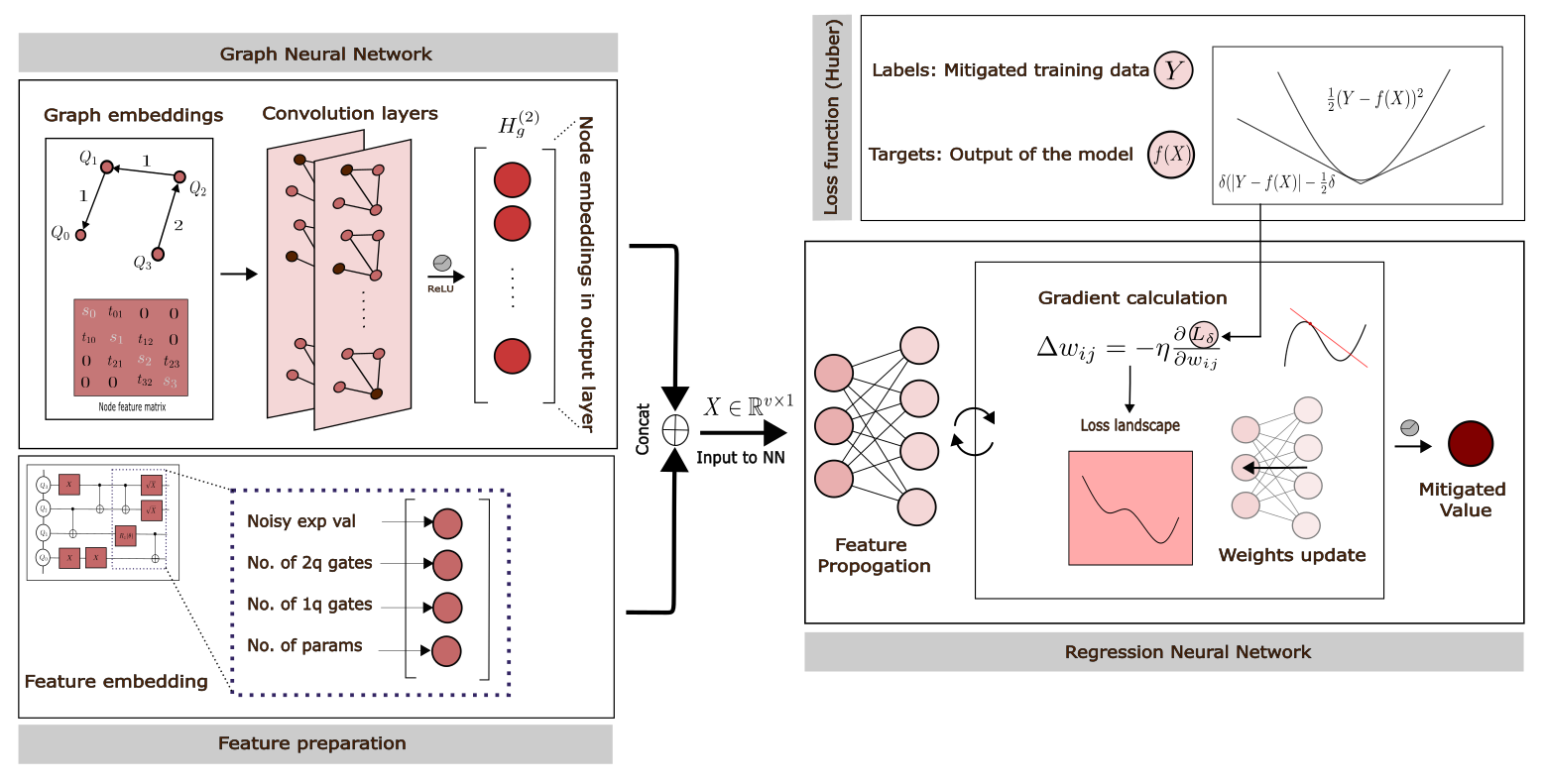}
    \caption{\textbf{GraphNetMitigator architecture: Output from GNN and features derived from the circuit are concatenated and serve as an input to the feed-forward neural network (regressor). The weights of the model are updated to minimize the Huber loss with hyperparameter $\delta$ which makes it robust against the outliers.}}
    \label{fig:gnn}
\end{figure*}

\subsection{Machine Learning Model Architecture}
We begin this section with the following disclaimer:
while in this particular work we generated an ansatz on the fly, the
ML model and the associated EM strategy that follow is independent 
of the way the ansatz is formed and in all aspects, is compatible
with any disentangled ansatz.
In this section, we give a detailed explanation of how the training data and training labels are generated, how the graph inputs are formulated which serve as an input to GNN, and the working of the entire model along with its complexity.

\subsubsection{Graph formulation}\label{sec: graph form}
As mentioned in the section \ref{sec: intro}, our training data consists of snippets of circuit ansatz consisting of one- and two-body fermionic excitations which only differ in terms of one and two-qubit gate applications. Therefore, graph representation is a natural modeling choice for such training data and GNNs can elegantly integrate such data structures. As mentioned in the section \ref{sec: tr data}, we derive the graph by transpiling the circuit adhering to the qubit-connectivity of the real quantum device having $n$ number of qubits. Graphs contain two types of information that we may wish to utilize for making predictions: nodes and edge connectivity. The node information can be embedded in the node feature matrix $X_g$. Expressing the graph's connectivity is not that straightforward. Using an adjacency matrix $A$ might be the most straightforward option, as it can be easily converted into a tensor format.  Here we explain how the graph and its corresponding attributes are constructed.

The graph $G(V, E)$ is represented as follows: The nodes $\{v_i\}$ correspond to the number of qubits $n$ in the device, and directed edges $e_{ij} \in E$ symbolize the CNOT connectivity between the $v_i$-th (parent node) and $v_j$-th (target node). The edge weights represent the frequency of identical edges. Subsequently, the matrix $\Tilde{A} = A + I$ is created to incorporate self-loops ($I$ is the identity matrix). The degree matrix $\Tilde{D}$ is derived from $\Tilde{A}$ to account for next-neighboring nodes. $\Tilde{D} = \text{diag}(\Tilde{D}_1, ..., \Tilde{D}_n)$ with $\Tilde{D}_{i} = \sum_j \Tilde{A}_{ij}$. $S = \Tilde{D}^{-\frac{1}{2}}\Tilde{A}\Tilde{D}^{-\frac{1}{2}}$ normalizes the adjacency matrix to incorporate information from neighboring nodes. 


The feature matrix $X_g \in \mathbb{R}^{n \times n}$ (g stands for GNN) of the nodes is then constructed. The diagonal elements $x_{ii}^g$ encode information about single-qubit gate errors specific to each qubit in the hardware used in the circuit. This information encompasses all single-qubit errors associated with the native gates. The off-diagonal elements $\{x_{ij}^g, i \neq j\}$ portray errors in two-qubit CNOT gates concerning the coupling map. The coupling map is a reflection of the physical connectivity of qubits in the hardware. The adjacency matrix $A \in \mathbb{R}^{n \times n}$ represents  CNOT connectivity with $A_{ij}$ representing the edge weight between two nodes $v_i$ and $v_j$. For more details, refer to Fig. \ref{fig: gnn embd}. The adjacency matrix representation suffers from drawbacks like redundancy and memory issues. One elegant and compact way to represent such sparse matrices is the adjacency list. It represents the connectivity of $k$-th edge $e_{mn}$ as a tuple $(m,n)$ in the $k$-th entry of the list. For instance, the adjacency list of the graph in Fig. \ref{fig: gnn embd} is given as $[(1, 0), (2, 1), (3, 2)]$.

\subsubsection{GraphNetMitigator Architecture and Training Details}

In this section, we describe the GNM architecture developed for error mitigation. We use a Graph Convolutional Network (GCN) to learn the features of the graph where each node corresponds to an $n$ dimensional embedding. Once we get the feature matrix $X_g$ and adjacency list of the graph, this input is passed to the GCN which transforms with the following propagation rule for $l$-th layer: 
\begin{equation}
    H_g^{( l+1)} = \sigma\left(SH_g^{(l)}W_g^{(l)}\right)
\end{equation}
 Where $W_g^{(l)} \in \mathbb{R}^{f_l \times f_{l+1}}$ is a feature transformation/weight matrix which maps the node features of size $f_l$ to size $f_{l+1}$. The GCN contains two convolutional layers. The first layer has $f_0 = n$ input channels reflecting the number of qubits present in the hardware and $f_1 = k$ output channels and the second layer has $f_2 = k$ input channels and $f_3 = 1$ output channel (This flow indicates the node feature transformation from $n \to k \to 1$ dimension ). When the first layer is applied to the input $X_g = H_g^{(0)}$, it transforms in a way with the weight matrix $W_g^{(0)} \in \mathbb{R}^{n \times k}$ as shown below. A ReLU activation function $\sigma(x) = \max(0, x)$ is applied on top of it. 
\begin{eqnarray}
H_g^{(1)}
   & = &  \sigma\left(SH_g^{(0)}W_g^{(0)}\right) \\ \nonumber 
   & = &  \sigma \left(\begin{bmatrix}
        h_{11}^{(1)} & ... & h_{1k}^{(1)} \\
        h_{21}^{(1)} & ... & h_{2k}^{(1)} \\
        \vdots & \ddots & \vdots\\
        h_{n}^{(1)} & ... & h_{nk}^{(1)} \\
        \end{bmatrix}_g \right) 
    =   \begin{bmatrix}
        H_{1}^{(1)} \\
        H_{2}^{(1)} \\
        \vdots \\
        H_{n}^{(1)} \\
        \end{bmatrix}_g
\end{eqnarray}
Where $h_{ij}^{(1)} = \sum_{m =1}^n x_{im}^g w_{mj}^g$. This represents the node representations after the first layer of the GCN. 
After passing through the second layer with a weight matrix of $W_g^{(1)} \in \mathbb{R}^{k \times 1}$ we get, 
\[
H_g^{(2)} = \begin{bmatrix}
H_{1}^{(2)} \\
H_{2}^{(2)} \\
\vdots \\
H_{n}^{(2)} \\
\end{bmatrix}_g = 
\begin{bmatrix}
h_{11}^{(2)} \\
h_{21}^{(2)}  \\
\vdots \\
h_{n1}^{(2)}  \\
\end{bmatrix}_g
\]
Where $h_{ij}^{(2)} = \sum_{m =1}^k x_{im}^g w_{mj}^g$. $H_g^{(2)}$ represents the node representations after the second layer of the GCN. This transformation corresponds to one-dimensional node embedding and serves as an input to our regressor along with additional features. 
 \begin{center}
    \begin{figure*}[!hbt]
        \centering
        \begin{tabular}{ccc}
    \includegraphics[scale=0.6]{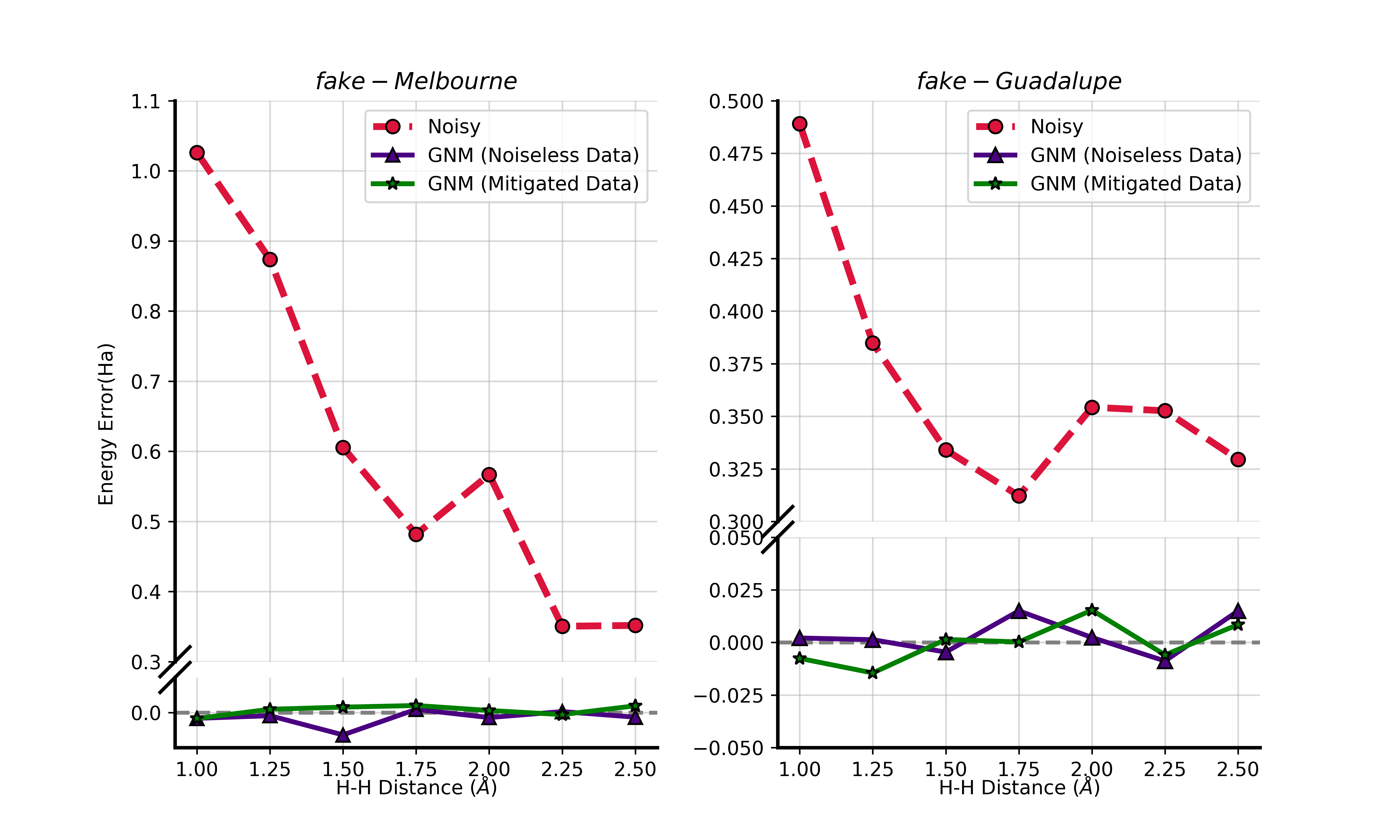}  
    \end{tabular}
         \caption{\textbf{The figure depicts the deviation of mitigated energy expectation value for the $H_4$ molecule from the ideal value in two different training scenarios. The purple plot represents the deviation of mitigated energy when ideal expectation values are used as targets in training data and the green plot represents the deviation of mitigated energy when sequential error mitigated expectation values are used as the targets for the training set Mitigated results on (left) \textit{fake-Melbourne} machine and (right) \textit{fake-Guadalupe} machine.}}
         \label{Fig:H4}
    \end{figure*}
 \end{center}
We construct the rest of the features from the following data set: \textit{noisy expectation values learnt over the snippet circuits, number of two-qubit gates, number of single-qubit gates, number of variational parameters (number of single and double excitation operators)}. We construct a feature vector $X_r \in \mathbb{R}^{l \times 1}$ ($r$ stands for regression) where $l$ is the number of features in regressor.
$H_g^{(2)}$ is then concatenated with $X_r$ to get final feature vector $X \in \mathbb{R}^{v \times 1}$  where $v = n + l$.  The quantity $X = (x_1, x_2, ..., x_{v})^T$ serves as 
an input to the feed-forward neural network with one hidden layer of $k_r$ = 64 nodes 
which is used for the regression task to mitigate the noisy value. First, $X$ 
transforms as follows: 
\begin{equation}
    H^{(1)} = \sigma \left( X \cdot W^{(0)} + b^{(0)} \right)
    \label{eq: layer1 reg}
\end{equation}
 $W^{(0)}$ and $b^{(0)}$ are weights and biases applied to the input layer with ReLU activation. Hidden layer $H^{(1)}$ also transforms in a similar manner giving the mitigated expectation values $f(X)$ as output:
 \begin{equation}
    f(X) = H^{(2)} = \sigma \left( H^{(1)} \cdot W^{(1)} + b^{(1)} \right)
    \label{eq: layer2 reg}
\end{equation}
The loss function used to calculate the loss is \textit{Huber loss}. The Huber loss function is often used in regression problems as a combination of the mean squared error and mean absolute error which is defined as:
\begin{center}
    \begin{figure*}[!hbt]
        \centering
        \begin{tabular}{ccc}
    \includegraphics[scale=0.6]{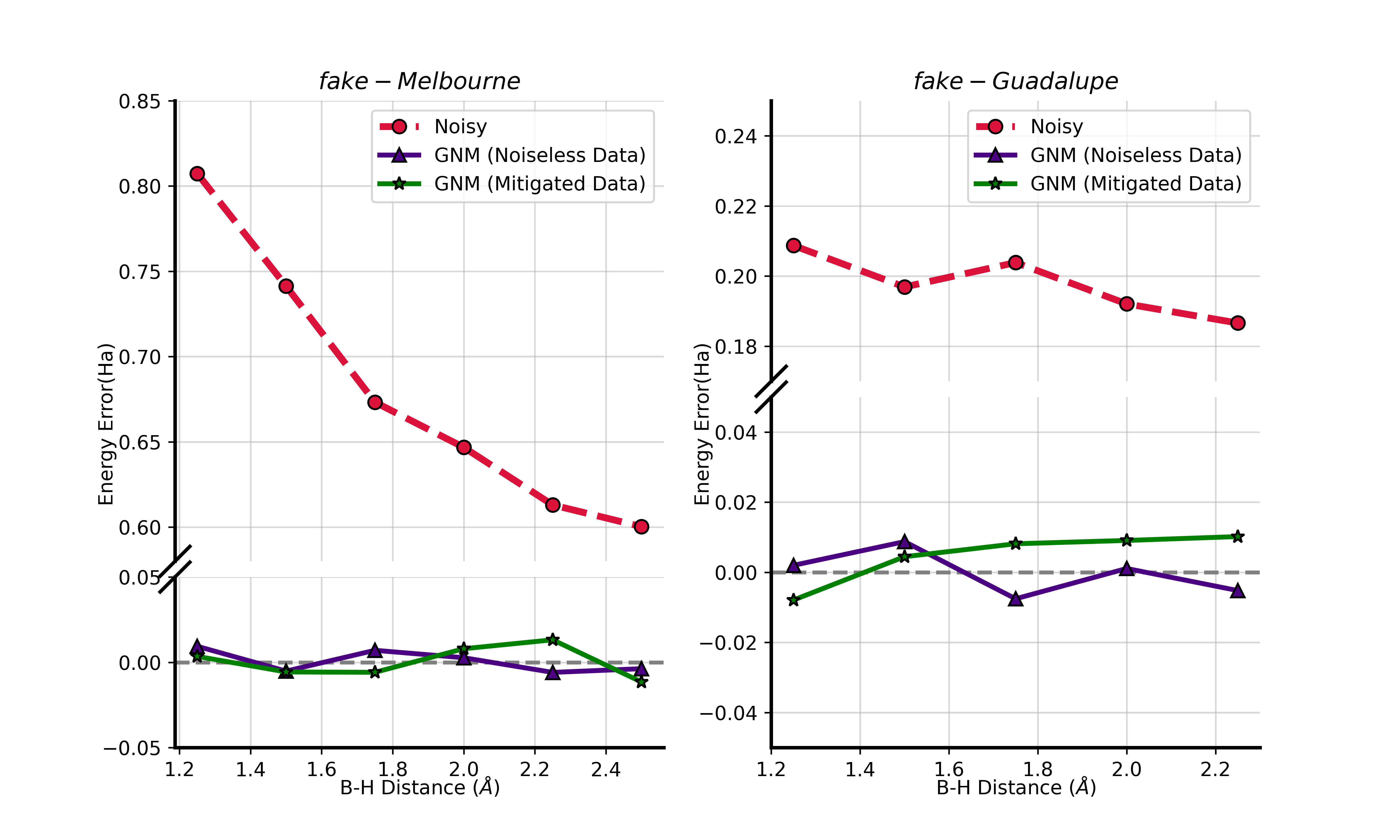}  
    \end{tabular}
         \caption{\textbf{The figure depicts the deviation of mitigated energy expectation value for the $BH$ molecule from the ideal value in two different training scenarios. The purple plot represents the deviation of mitigated energy when ideal expectation values are used as targets in training data and the green plot represents the deviation of mitigated energy when sequential error mitigated expectation values are used as the targets for the training set Mitigated results on (left) \textit{fake-Melbourne} machine and (right) \textit{fake-Guadalupe} machine.}}
         \label{Fig:BH}
    \end{figure*}
 \end{center}
\begin{equation}
    L_{\delta}(Y, f(X)) = \begin{cases}
        \frac{1}{2}(Y - f(X))^2, & \text{if } |Y - f(X)| \leq \delta, \\
        \delta(|Y - f(X)| - \frac{1}{2}\delta), & \text{otherwise}.
    \end{cases}
\end{equation}
Where $f(X)$ is the mitigated prediction and $Y$ is the ideal/SREM value (from the snippets)
used as a target to calculate the loss. Here $\delta$ is a hyper-parameter determining where 
the loss transitions from quadratic to linear behavior. In our case, $\delta$ is defined by 
analyzing the mean square error between ideal/SREM values of the snippets and the
corresponding noisy expectation values $\delta E_\alpha^{SREM} = |E^{SREM}_{\alpha} - E_\alpha|^2$ and 
$\delta E_\alpha^{ideal} = |E^{ideal}_\alpha - E_\alpha|^2$ for $\alpha$ indicating 1, 2, 3-parameter snippet circuits from the training set to handle the outliers. The choice of the loss function is based on the fact that with a given circuit, sampling cost scales exponentially with respect to its depth implying eventual intractability in mitigating associated errors\cite{takagi2022fundamental}. Giving less weightage to such non-shallow circuit expectation values which show large deviation $\delta E_\alpha^{SREM}$ when SREM values are used as labels or $\delta E_\alpha^{ideal}$ when ideal values are used as labels ensures the proper handling of outliers.

\subsubsection{Complexity analysis of the model}
\textbf{Time and Space complexity of GCN:}
For matrices $A \in \mathbb{R}^{a \times b}$ and $B \in \mathbb{R}^{b \times c}$, matrix multiplication $AB$ has a time complexity of $O(abc)$.
We use pytorch geometric to implement the GCN. PyTorch Geometric first computes the matrix multiplication of $Z_g$ of $H^{(l)}_g$ and $W^{(l)}_g$. To compute $S$, they use coordinate form edge index matrix of size $2 \times |E|$ and store the normalization coefficient in a separate matrix called \textit{norm}. The computation of $ SZ_g$ is what involves the so-called "message passing" and aggregation. So it's done in a two-step process.
The first step is a dense matrix multiplication between matrices of size $n \times n$ and $n \times k$. So the time complexity becomes $O(n\cdot n \cdot k)$. The second step matrix multiplication of $n \times n$ and $n \times k$ will have a similar time complexity. In practice, it is computed using a sparse operator, such as the PyTorch scatter
function. The activation is simply an element-wise function with a $O(n)$ complexity. Considering the neighborhood aggregation of each node, over $l$ layers the total time complexity becomes $O(ln^2k + l |E|k )$. Backward pass also has a similar time complexity of $O(ln^2k + l |E|k )$. More details can be found in  \cite{blakely2021time}. The overall time and space complexity of the GCN model is described in the table below: 

\textbf{Time and Space complexity of Regressor:} 
The regressor uses a simple feed-forward neural network given by the Eq. \ref{eq: layer1 reg} and Eq. \ref{eq: layer2 reg}. The matrix multiplication of $X$ and $W$ has a time complexity of $O(v \cdot 1 \cdot k_r)$ = $O(nk_r)$. Adding the activation layer leads to the total time complexity of $O(l_rnk_r + l_rn)$ $\approx$ $O(l_rnk_r)$ for $l_r$ layers. 

\begin{table}[t]
\centering
\begin{tabular}{p{4 cm}p{3 cm}}
\hline
\hline
\textbf{Algorithm} &  \textbf{Time Complexity} \\
\hline
GCN forward pass & $O(ln^2k + l |E|k )$ \\
GCN backward pass & $O(ln^2k + l |E|k )$ \\
Regressor forward pass & $O(l_rnk_r  )$ \\
Regressor backward pass & $O(l_rnk_r  )$ \\
\hline
\hline
\end{tabular}
\caption{\textbf{Time and space complexity analysis of forward and backward pass.}}
\label{table}
\end{table}

\section{Results and Discussions:}
To demonstrate the working of the GNM (shown in Fig. \ref{fig:gnn}), we test it on two different molecules and two different devices with diverse noise profiles. 
Since the model is not based on the variational principle, the estimated expectation value can go below the ideal value. With the advent of modern gate error suppression techniques, such as dynamic decoupling
and gate twirling, fluctuations in measurement can be greatly suppressed. In a 
simulation protocol, this can be approximated by using a fixed \textit{seed} value to represent an averaged sampling statistics which we have chosen to use throughout 
this work. In this architecture, we train an ML model with circuits composed of 
single and double excitations to predict noiseless expectation values using noisy 
ones. The GNM model is developed using PyTorch\cite{pytorch} and PyTorch Geometric. For GNN, 
we use Graph Convolution Networks. The details of graph encoding and feature 
construction are found in the Methodology section. In the training data, labels 
are generated sequentially using the reference error mitigation
strategy which uses VQE as a subroutine and is implemented using Qiskit-
Nature\cite{qiskit_nature}. The one- and two-body integrals are imported
from PySCF\cite{pyscf}. The second-quantized fermionic operators are 
transformed into qubit operators by using Jordan-Wigner 
encoding\cite{JWT}. All the noise models used are derived from fake IBM 
Quantum hardware. We mainly use noise models imported from fake backends 
of two quantum devices: \textit{ibmq Melbourne} and \textit{ibmq 
Guadalupe} having 14 and 16 qubits respectively. The model is trained 
separately for each bond length using the Adam optimizer with 100 
epochs. Due to a limited number of data in the training set, batch size 
is taken as 1.  

To illustrate the functionality of the GNM, our initial testing 
focuses on the $H_4$ molecule where we symmetrically stretch all the $H-H$ 
bonds. In STO-3G basis $H_4$ has 4 electrons in 8 spin-orbitals. From the findings 
depicted in Fig. \ref{Fig:H4}, it becomes evident that the optimized energy 
difference for full circuit between the noisy and corresponding noiseless 
scenarios ranges from approximately 1 $E_h$ to 0.32 $E_h$ on the \textit{ibmq-Melbourne}
machine and from 0.48 $E_h$ to 0.3 $E_h$ on the \textit{ibmq-Guadalupe} machine. 
However, when considering the mitigated energy compared to the noiseless energy, 
the difference is confined within 12 $mE_h$ on the \textit{ibmq-Melbourne} machine 
when the GNM model is trained with either the noiseless (simulator generated) or
SREM training datasets of the snippets. Similarly, on the \textit{ibmq-Guadalupe} 
machines, the maximum observed error for GNM mitigated energy reaches around 20 $mE_h$ 
for both the noiseless and SREM training datasets generated from the snippets.

The stretching of the $B-H$ bond represents one of the most challenging test cases for 
NISQ implementations. In STO-3G basis with frozen 1s core orbital of $B$, it turns out 
to be a system of 4 electrons in 10 spin-orbitals. 
Observing Figure \ref{Fig:BH}, it becomes evident that the energy error between the 
noisy and noiseless scenarios for full circuit fall within the range of 0.82 to 0.59 $E_h$ 
on \textit{fake-ibmq-Melbourne} device, and within 0.21 to 0.19 $E_h$ on the \textit{fake-ibmq-
Guadalupe} device. In contrast, the GNM model predicts the mitigated energies which 
are within 20 $mE_h$ and 18 $mE_h$ (at max over the potential energy surface, with respect 
to the corresponding noiseless values) on \textit{fake-ibmq-Melbourne} and 
\textit{fake-ibmq-Guadalupe} machine respectively, irrespective of the training dataset
used for the snippets.

\begin{center}
    \begin{figure}[!t]
        \centering
        \includegraphics[width=1\linewidth]{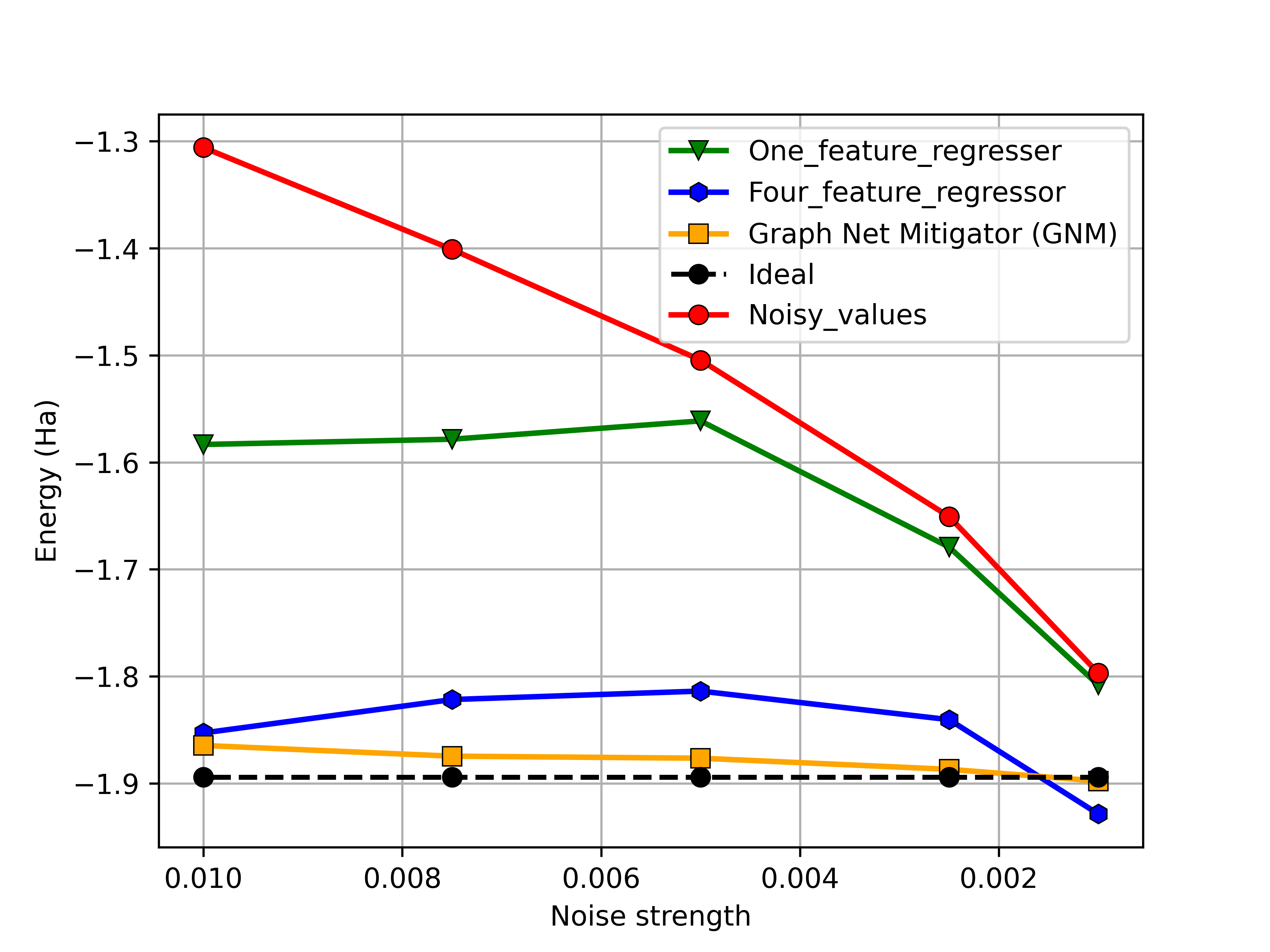}
        \caption{\textbf{Noisy and Mitigated energy expectation values are plotted at
        different two-qubit depolarizing noise strength. Three different models (regression model with only one feature, regression model with additional circuit features, and GraphNetMitigator model) are trained under these scenarios.}}  
        \label{fig: varnoise}
    \end{figure}
\end{center}

We also tested the robustness of our model against varying noise levels for $H_4$ at
2\AA. For this particular analysis, we specifically accounted for the depolarizing 
noise (which is one of the primary sources of noise in NISQ hardware) with noise
strength varying from \{0.0001, 0.001\} to \{0.001, 0.01\} for one-qubit and 
two-qubit gates, respectively. Here the training labels are generated via the 
SREM mitigated values.
The results are shown in Fig. \ref{fig: varnoise}. 
We test for different 
variations of machine learning models, from only one-feature regression to GNM 
and in all occasions, GNM predicts the most accurate energy, establishing
its robustness across a spectrum of noise strengths. It is particularly interesting to
note that with the decrease in noise strength, as expected with the improvement of 
the hardware quality, GNM predicted values linearly approach the corresponding 
'exact' value, demonstrating its ability to handle the spectrum of noise strength.
without overestimating or underestimating. Despite its good 
performance with some of the most dominant noise channels as discussed so far, 
its robustness is subject 
to further validation when the state preparation and measurement (SPAM) errors, 
without sampling noise suppression models\cite{DD1,Twirling}, are accounted for. In principle, 
a separate SPAM error mitigation strategy may be complemented as an
additional mitigation layer for better state preparation. However,
the current graph network-based model gives
us a practical tool to mitigate gate errors that otherwise
incur exponential overheads and deep mitigation circuits.

\section{Conclusions and Future Outlook}
In this work, we developed a learning-based error mitigation technique 
to accurately predict the ground state energy of molecular Hamiltonian. 
The developed GNM model successfully predicts the ground 
state energy of the many-electron Hamiltonian without the need for any 
additional mitigation circuits and bypasses exponential overhead. 
GNM outperforms
current EM methods like digital-ZNE and PEC which fail at larger 
and complex circuits. Unlike other NN-based techniques which often use 
ideal expectation values generated on the simulator as labels, we
generate our labels, either by ideal values or by sequentially mitigating 
the reference error over much shallower snippet circuits, thus eliminating the 
need to access the ideal quantum computer. The data for training 
the model is generated on the fly while the ansatz is constructed
in a noisy quantum environment. GNM is
entirely general mitigation model, and does not have any specific dependence to the 
underlying ansatz. Our results consistently demonstrate the robustness 
and effectiveness of GNM across hardware connectivity and 
a spectrum of synthetic noise strengths, and thus GNM is 
expected to hold its ground when the qubit quality improves over the 
coming years.

One important future research direction is the generalization of the 
model across diverse and larger classes of Hamiltonians and ansatze. 
Testing the model while considering different noise 
channels with nonlinear sampling complexity is an important 
future direction towards establishing its robustness.
More importantly, feature selection can be improved using a different 
set of features altogether. The current method captures the device 
characteristics through a CNOT-connectivity graph adhering to the 
coupling map of the device and gate errors. An additional set of 
features like decoherence time of qubits and other device-specific 
parameters can be incorporated to further capture the noise profile 
accurately. 

\section*{Software}
The current model supports Qiskit SDK and is developed using Torch and Torch-Geometric. For mitigation purposes, we use \texttt{AerEstimator} to calculate the noisy values and subsequently mitigated SREM values. We provide detailed documentation and source code of training data generation through SREM and GraphNetMitigator model at \textcolor{blue}{\url{https://github.com/Next-di-mension/GraphNetErrorMitigator}}
\section*{Supplementary Material}
See supplementary material for the Sequential Reference Error Mitigation (SREM) strategy and 
the distribution of the all the noisy and REM mitigated energies for two representative 
molecules in two extreme geometries that were used for label generation.
\section*{Acknowledgements}
S.P. Thanks to the Industrial Research and Consultancy Center (IRCC), IIT Bombay. S.P. is also thankful to Ananyapam De (Claushtal University of Technology, Germany) for useful discussions on developing the ML model. D.M. acknowledges the Prime Ministers Research Fellowship (PMRF), Government of India for their research fellowships. R.M. acknowledges the financial support from the Anusandhan National Research Foundation 
(ANRF) (erstwhile SERB), Government of India (Grant Number: MTR/2023/001306).

\section*{Conflict of Interests}
The authors have no conflict of interest to disclose.

\section*{Data Availability}
The data that support the findings of this study are available from the corresponding author upon reasonable request.

\bibliography{ref}
\end{document}


\author{Srushti Patil}
\thanks{These two authors contributed equally.}
\affiliation{ NNF Quantum Computing Programme, \\ Niels Bohr Institute,
University of Copenhagen, Jagtvej \\ 155 A, Copenhagen N DK, 2200 \\ Copenhagen, Denmark}

\author{Dibyendu Mondal}
\thanks{These two authors contributed equally.}
\affiliation{ Department of Chemistry,  \\ Indian Institute of Technology Bombay, \\ Powai, Mumbai 400076, India}

\author{Rahul Maitra}
\email{rmaitra@chem.iitb.ac.in}
\affiliation{ Department of Chemistry,  \\ Indian Institute of Technology Bombay, \\ Powai, Mumbai 400076, India}
\affiliation{Centre of Excellence in Quantum Information, Computing, Science \& Technology, \\ Indian Institute of Technology Bombay, \\ Powai, Mumbai 400076, India}

\title{Supplementary Materials: Machine Learning Approach towards Quantum Error Mitigation for Accurate Molecular Energetics}
\maketitle

\section{Sequential Reference Error Mitigation (SREM) Strategy for Snippet Circuits}
In the supplementary materials, we discuss the Sequential Reference Error 
Mitigation (SREM) strategy, which is designed to 
mitigate reference errors in small snippet circuits for training label generation. 
We consider a prototype circuit,
($\ket{\psi}=e^{\tau_L}e^{\tau_K}e^{\tau_J}\ket{\phi_{HF}}$) which consists 
of three excitation operators $(\tau_L, \tau_K, \tau_J)$, which is of maximum length
our model is trained over.

We begin sequentially, starting with the conventional reference error mitigation 
strategy\cite{REM} for a one-parameter circuit, $\ket{\psi}=e^{\tau_J}\ket{\phi_{HF}}$, 
where the excitation operator $\tau_J$ acts on the reference Hartree-Fock (HF) state, 
whose exact energy value ($E_{HF}$) is known from classical computation. Ideally, 
in the absence of noise, setting the associated parameter $\theta_J=0$ in the 
circuit should yield the HF energy ($E_{HF}$). However, due to hardware noise, 
the measured reference energy for the one-parameter circuit 
$E_J^0=\bra{\phi_{HF}}e^{-\tau_J(\theta_J=0)}He^{\tau_J(\theta_J=0)}\ket{\phi_{HF}}$
significantly deviates from $E_{HF}$. The characteristic error in calculating the 
reference energy due to noise in the one-parameter circuit is given by
\begin{eqnarray}
    \Delta_J^{ref} = E_{HF} - E_J^0
\end{eqnarray}
Next, we variationally obtain the optimized energy $E_J$ as a function of 
$\theta_J$, starting from an initial value of zero:
\begin{eqnarray}
    E_J = \bra{\phi_{HF}}e^{-\tau_J(\theta_J)}He^{\tau_J(\theta_J)}\ket{\phi_{HF}}=E_J^{noisy}
\end{eqnarray}
This optimized energy $E_J, \forall J$ is the noisy energy values for one parameter 
circuits and serves as one of the training data for the GNM model. Notably, in 
the case of an extremely shallow circuit with a single variational parameter, 
the differences $(E_J- E_J^0)$ and $(E_J^{noiseless} - E_{HF})$ are approximately 
equal. Therefore, for the one-parameter circuit, the mitigated energy is given by
\begin{eqnarray}
    E_J^{EM} = E_J + \Delta_J^{ref}
\end{eqnarray}
The mitigated energy values $E_J^{EM}, \forall J$ are then utilized as noiseless 
training data for the GNM model and serve as the noiseless reference energy for 
subsequent steps in the SREM procedure.

We now proceed with the error mitigation for a two-parameter circuits,
($\ket{\psi}=e^{\tau_K}e^{\tau_J}\ket{\phi_{HF}}$) which involves the excitation operators
$\tau_K$ and $\tau_J$. With $\theta_J$ set to its optimized value $\Bar{\theta}_J$, 
obtained from the previous one-parameter circuit optimization, we construct the 
energy functional for the two-parameter system and estimate the energy:
\begin{eqnarray}
E_{JK} = \bra{\phi_{HF}}e^{-\tau_J(\Bar{\theta}_J)}e^{-\tau_K(\theta_K)} 
    \hat{H}e^{\tau_K(\theta_K)}e^{\tau_J(\Bar{\theta}_J)}\ket{\phi_{HF}}. 
\end{eqnarray}
Here, $\theta_K$ is taken as the sole variational parameter to be optimized in this step.
This setup is equivalent to the single-parameter energy optimization, where the state
$e^{\tau_J(\Bar{\theta}_J)}\ket{\phi_{HF}}$ is treated as the reference, 
with the ‘mitigated’ reference energy $E_J^{EM}$ obtained from the previous step.
In an ideal environment, setting $\theta_K = 0$ should restore $E_J^{EM}$, given 
that $\theta_J = \Bar{\theta_J}$. 
However, noise introduces a deviation in the reference energy, which is quantified as:
\begin{eqnarray}
    \Delta_{JK}^{ref} = 
    E_J^{EM} - 
    \bra{\phi_{HF}}e^{-\tau_J(\Bar{\theta}_J)}e^{-\tau_K(\theta_K =0)}
    \hat{H}e^{\tau_K(\theta_K =0 )}e^{\tau_J(\Bar{\theta}_J)}\ket{\phi_{HF}}  
    = E_J^{EM} - (E_K^0)_J \;\;\;\;\;\;\;\;\;\;\;\;\;\;\;\;\;\;\;\;\;\;\;\;\;\;\;\;\;\;\;
\end{eqnarray}  
Where $(...)_J$ indicates that the reference state is taken as $e^{\tau_J(\Bar{\theta}_J)}\ket{\phi_{HF}}$.
Similar to the one-parameter case, we obtain the mitigated energy estimate for the two-parameter 
circuit:
\begin{equation}
    E_{JK}^{EM} = E_{JK} + \Delta_{JK}^{ref}.
\end{equation}
It is important to note that $E_{JK}$ is obtained by optimizing $\theta_K$
while keeping $\theta_J$ fixed at its previously optimized value $\Bar{\theta}_J$. As a result, the full correlation energy of the two-parameter circuit is not fully captured in $E_{JK}$.
To address this, a full two-parameter optimization is performed, starting from the previously 
optimized values:
\begin{eqnarray}
    E_{JK}' =\bra{\phi_{HF}}e^{-\tau_J(\theta_J)}e^{-\tau_K(\theta_K)} 
    \hat{H}e^{\tau_K(\theta_K)}e^{\tau_J(\theta_J)}\ket{\phi_{HF}} 
\end{eqnarray}
Here, both $\theta_K$ and $\theta_J$ are optimized, yielding new optimized values
$\Bar{\theta}_K^{'}$ and $\Bar{\theta}_J^{'}$ respectively. 
The final SREM-mitigated energy for the two-parameter circuit is then given by:
\begin{eqnarray}
    E_{JK}^{SREM} & = & E_{JK}^{EM} + \left( E_{JK}' -  E_{JK}\right) 
    =  E_{JK} + \Delta_{JK}^{ref} +  E_{JK}' -  E_{JK} 
    =   E_{JK}' +   E_{J}^{EM} -   (E_{K}^0)_J
    \label{eq: 2p srem}
\end{eqnarray}
The fully optimized two-parameter circuit, $\ket{\psi}=e^{\tau_K(\Bar{\theta}_K^{'})}e^{\tau_J(\Bar{\theta}_J^{'})}\ket{\phi_{HF}}$ and the corresponding SREM energy $E_{JK}^{SREM}$ serve as the reference state and noiseless reference energy, respectively, for mitigating errors in the three-parameter circuit.
For the two-parameter circuit, it is important to note that $E_{JK}^{SREM}$ serves as the noiseless energy data, while $E_{JK}^{noisy}$ is used as the noisy energy data for training the
GNM model. This noisy energy $E_{JK}^{noisy}$ is obtained by optimizing both parameters starting from an initial value of zero:
\begin{eqnarray}
    E_{JK}^{noisy} =\bra{\phi_{HF}}e^{-\tau_J(\theta_J)}e^{-\tau_K(\theta_K)} 
    \hat{H}e^{\tau_K(\theta_K)}e^{\tau_J(\theta_J)}\ket{\phi_{HF}} (initial-point = [0,0])
\end{eqnarray}

\begin{figure*}[t]
    \centering
    \includegraphics[width= 14cm, height=10cm]{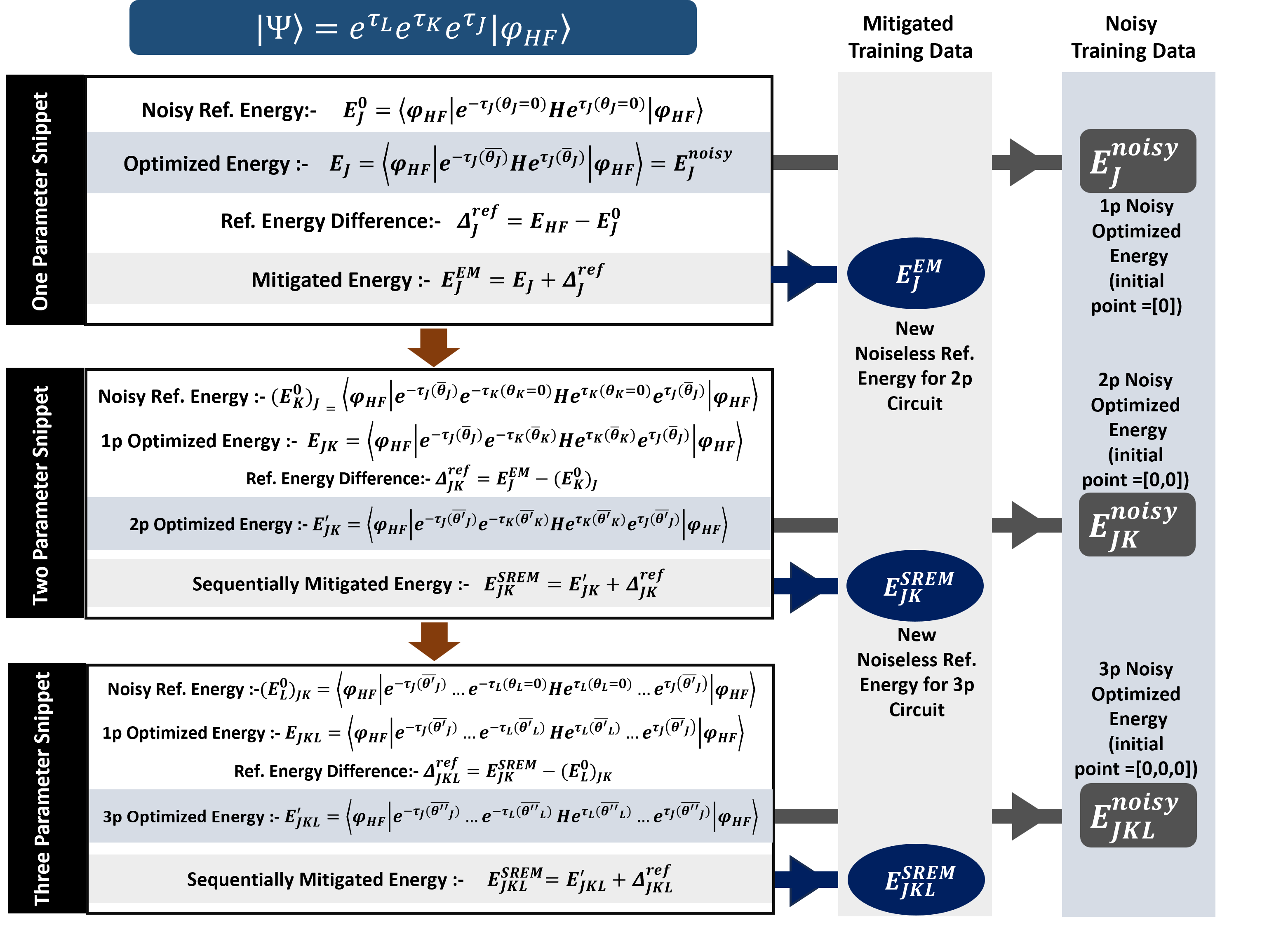}
    \caption{\textbf{Schematic representation for generating training data sets: SREM and Noisy energy for one, two, and three-parameter circuits.}}
    \label{fig:SREM}
\end{figure*}

Next, we consider the mitigation procedure for circuits parameterized by three operators, where the ansatz is defined as $\ket{\psi}=e^{\tau_L}e^{\tau_K}e^{\tau_J}\ket{\phi_{HF}}$.
Following a similar strategy as before, we construct a three-parameter energy functional. Here, we optimize $\theta_L$ as the sole variational parameter while fixing the other parameters 
$\theta_J$ and $\theta_K$ at their previously optimized values,
$\Bar{\theta}_J^{'}$ and  $\Bar{\theta}_K^{'}$ respectively. The energy evaluated under noise with this single-parameter optimization for the operator $\tau_L$ is given by
\begin{eqnarray}
   E_{JKL} = \bra{\phi_{HF}}e^{-\tau_J(\Bar{\theta}_J^{'})}e^{-\tau_K(\Bar{\theta}_K^{'})}e^{-\tau_L(\theta_L)}   \hat{H}e^{\tau_L(\theta_L)}e^{\tau_K(\Bar{\theta}_K^{'})}e^{\tau_J(\Bar{\theta}_J^{'})}\ket{\phi_{HF}}
\end{eqnarray}
Similarly, as discussed above the noisy reference energy for the three-parameter circuit is defined as
\begin{eqnarray}
    (E_{L}^0)_{JK} = \bra{\phi_{HF}}e^{-\tau_J(\Bar{\theta}_J^{'})}e^{-\tau_K(\Bar{\theta}_K^{'})}e^{-\tau_L(\theta_L=0)}
\hat{H}e^{\tau_L(\theta_L=0)}e^{\tau_K(\Bar{\theta}_K^{'})}e^{\tau_J(\Bar{\theta}_J^{'})}\ket{\phi_{HF}}
\end{eqnarray}
where, $(...)_{JK}$ suggests that the reference state is taken as $e^{\tau_J(\Bar{\theta}_J^{'})}e^{\tau_K(\Bar{\theta}_K^{'})}\ket{\phi_{HF}}$. 
Therefore, the deviation in the reference state energy for the three-parameter circuit is expressed as
\begin{eqnarray}
    \Delta_{JKL}^{ref}  
    = E_{JK}^{SREM} - (E_{L}^0)_{JK} 
\end{eqnarray}
the mitigated energy with respect to the noise for the one-parameter optimization of
$\tau_L$, becomes
\begin{equation}
    E_{JKL}^{EM} = E_{JKL} + \Delta_{JKL}^{ref}.
\end{equation}
Finally, all three parameters are released as free variational parameters. The energy functional is then re-optimized, starting from their previously determined optimal values, yielding
\begin{eqnarray}
   E_{JKL}^{'} = \bra{\phi_{HF}}e^{-\tau_J(\theta_J)}e^{-\tau_K(\theta_K)}e^{-\tau_L(\theta_L)}   \hat{H}e^{\tau_L(\theta_L)}e^{\tau_K(\theta_K)}e^{\tau_J(\theta_J)}\ket{\phi_{HF}}
\end{eqnarray}
Thus, the final SREM energy for the three-parameter ansatz is given by
\begin{eqnarray}
    E^{SREM}_{JKL}& = &E^{EM}_{JKL}+(E_{JKL}^{'} -E_{JKL}) \\ \nonumber
    & = & E_{JKL} +\Delta_{JKL}^{ref} + (E_{JKL}^{'}-E_{JKL}) \\ \nonumber
    &=& (E^{SREM}_{JK}- (E_{L}^0)_{JK}) +(E_{JKL}^{'})
\end{eqnarray}
This completes the mitigation process for the three-parameter circuit and $E^{SREM}_{JKL}$
is used as noiseless energy for training in GNM model. The noisy energy $E_{JKL}^{noisy}$
for three parameter circuit which is used as training data for GNM model is obtained by
optimizing all three parameters $\theta_J$, $\theta_K$ and $\theta_L$ starting from
the initial point zero.
\begin{eqnarray}
   E_{JKL}^{noisy} = \bra{\phi_{HF}}e^{-\tau_J(\theta_J)}e^{-\tau_K(\theta_K)}e^{-\tau_L(\theta_L)}   \hat{H}e^{\tau_L(\theta_L)}e^{\tau_K(\theta_K)}e^{\tau_J(\theta_J)}\ket{\phi_{HF}}
   (initial-point=[0,0,0])
\end{eqnarray}

The detailed algorithm for the SREM technique is shown in Fig. \ref{fig:SREM}.

\begin{center}
    \begin{figure*}[!hbt]
        \centering
        \begin{tabular}{ccc}
    \includegraphics[scale=0.6]{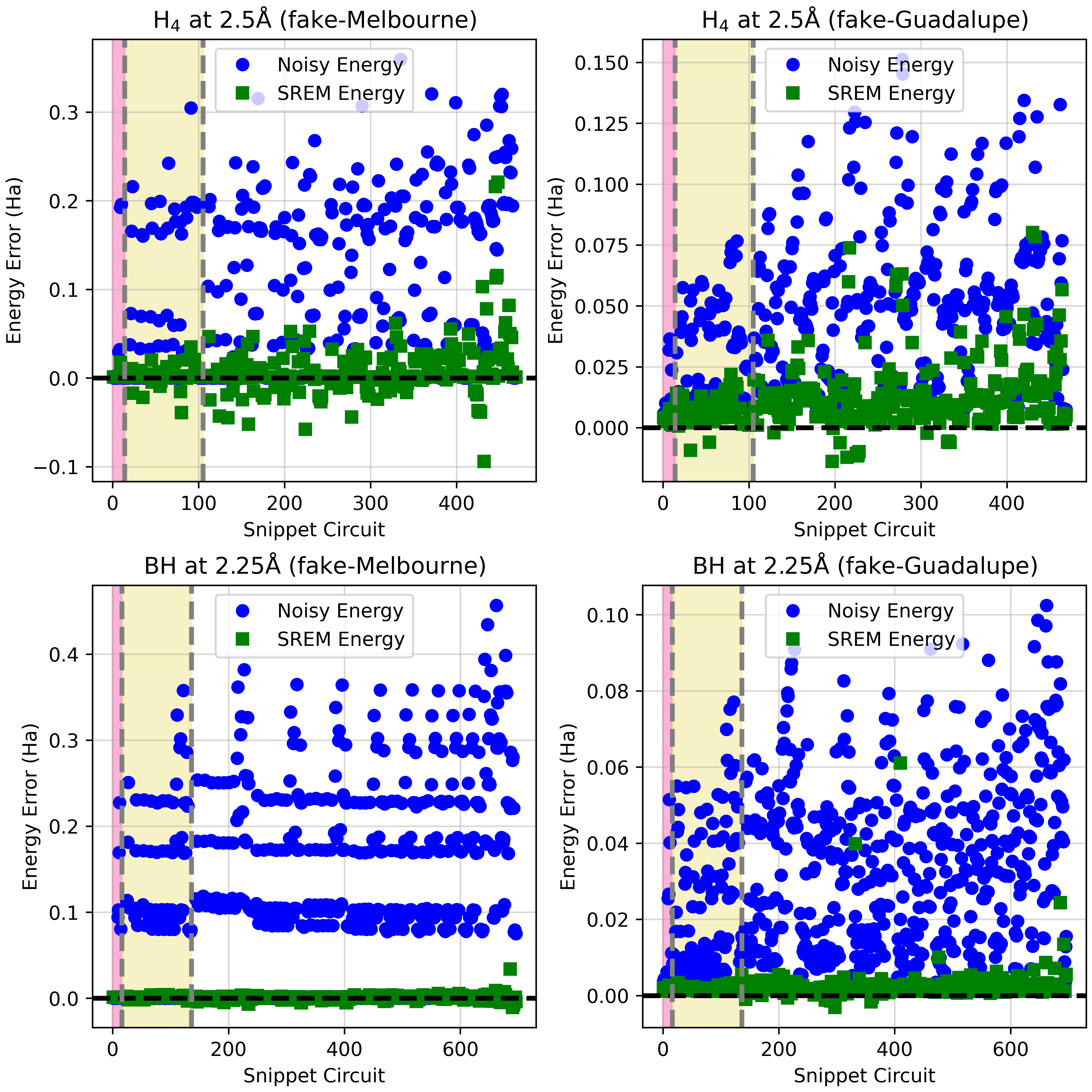}  
    \end{tabular}
         \caption{\textbf{The energy error relative to the noiseless value is plotted for both noisy and SREM energies across various snippet circuit combinations. The pink region corresponds to the one-parameter circuit, the light yellow region represents the two-parameter circuit, and the white region denotes the three-parameter circuit. The mitigated expectation values show larger spread as one moves from one to two to three parameter snippet circuits.}}
         \label{Fig:Training}
    \end{figure*}
 \end{center}
\bibliography{ref}